# High Resolution Mapping of Phase Behavior in a Ternary Lipid Mixture: Do Lipid-Raft Phase Boundaries Depend on Sample-Prep Procedure?

Jeffrey T. Buboltz, Charles Bwalya, Krystle Williams and Matthew Schutzer

*Department of Physics and Astronomy, Colgate University, Hamilton, NY 13346*

For some time now, we have been using a FRET-based strategy to make high-resolution studies of phase behavior in ternary lipid-raft membrane mixtures. Our FRET experiments can be carried out on ordinary, polydisperse multilamellar vesicle suspensions, so we are able to prepare our samples according to a procedure that was designed specifically to guard against artifactual phase separation. In some respects (i.e., the number and nature of two-phase regions observed), our phase diagrams are consistent with previously published reports. However, in other respects (i.e., overall size of miscibility gaps, phase boundary locations and their dependence on temperature) there are clear differences. Here we present FRET data taken in DOPC/DPPC/Cholesterol mixtures at 25.0, 35.0 and 45.0°C. Comparisons between our results and previously reported phase boundaries suggest that lipid-raft mixtures may be particularly susceptible to demixing effects during sample preparation.

## Introduction

In recent years, the phase behavior of "raft-like" membrane-lipid mixtures has been the subject of intense research,[1] with certain ternary mixtures receiving especial attention, as discussed in an excellent review by Veatch and Keller.[2] A number of different investigative techniques have been employed in these studies, but by far the most influential technique has been confocal fluorescence microscopy (CFM). Nevertheless, as a tool for investigating composition-dependent phase behavior, CFM suffers from two limitations. First, there is the tendency of mixture components to demix during GUV preparation.[2] And second, only a relatively small number of independently prepared samples can be characterized, due to the considerable time, effort and skill required by CFM.

We have recently described an experimental technique that was specifically developed for mapping composition-dependent phase behavior in membrane mixtures. This technique, which we call steady-state probe-partitioning FRET (SP-FRET), is not subject to the above limitations and is particularly well suited for the generation of high-resolution data sets. In our paper detailing the technique,[3] we demonstrated SP-FRET's sensitivity to the presence of coexisting membrane domains in the simplest possible context: a binary mixture with coexisting $L_\alpha$ and $L_\beta$ membrane phases at 20°C.

In this report, we have used SP-FRET to map three different regions of phase coexistence—at three different temperatures—in the ternary mixture DOPC/DPPC/Cholesterol. Our results convey two basic messages. First, they illustrate the effectiveness of SP-FRET as a general tool for mapping phase behavior in ternary mixtures. Second, they suggest that the apparent phase behavior of lipid-raft mixtures may be particularly sensitive to demixing artifacts, so that the choice of sample preparation technique should be given very careful consideration in any study of lipid-raft phase behavior.

## Experimental Section

**Chemicals.** DOPC, DPPC and cholesterol were purchased from Avanti Polar Lipids and purity was confirmed by thin layer chromatography on washed, activated silica gel plates as previously described.[4] Donor and acceptor probes, dehydroergosterol (DHE) and 3-3'-dioctadecyloxacarbocyanine (18:0-DiO), were from Sigma-Aldrich and Invitrogen, respectively. PIPES buffer and disodium EDTA were from Fluka Chemie AG. Aqueous buffer (2.5mM PIPES pH 7.0, 250mM KCl, 1mM EDTA) was prepared from 18 MΩ water (Barnstead E-Pure) and filtered through a 0.2 μm filter before use.

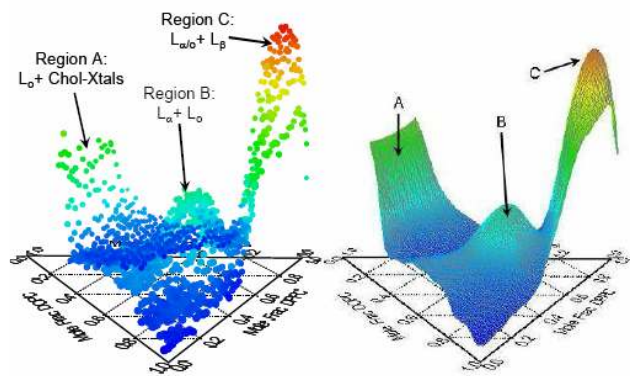

Fig. 1. SP-FRET measurements reveal clearly the locations of three different regimes of coexisting lipid phases in DOPC/DPPC/Chol MLV suspensions at 25°C. Changes in donor-excited acceptor fluorescence ($-\Delta(F_{DiO}^{DHEex})_{norm}$, see text) are plotted vs. lipid composition in triangular coordinates. In the scatter plot on the left, each data point corresponds to an independently prepared sample (1294 total). The right-hand plot shows a smooth surface fit to the same data. Probe mole fractions were fixed at $\chi_{DHE} = 3.0\times 10^{-3}$ and $\chi_{DiO} = 3.0\times 10^{-4}$ for all samples.

---

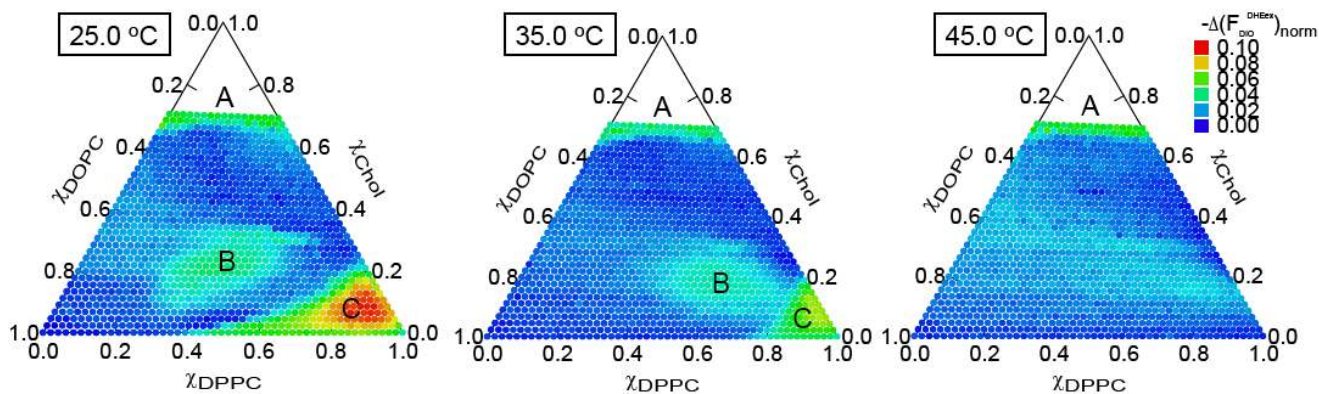

Fig. 2. Effect of temperature on phase behavior: 2D Mosaic Plots. Within each triangle, each data point represents an independently prepared sample (left panel is same data shown in Fig. 1). DHE-excited 18:0-DiO fluorescence was measured at each of the three temperatures shown, allowing two days for equilibration. As temperature changes from 25°C to 35°C, coexistence regions B and C clearly shift toward higher DPPC content, and by 45°C neither region can be seen to persist. The same twenty-degree change has no observable effect on the cholesterol crystalline phase boundary defining region A.

**Sample preparation.** Specified sample compositions ($4.5x10^{-7}$ moles total lipid per sample) were prepared in 13 x 100 mm screw cap tubes by combining appropriate volumes of chloroform-based lipid and probe stock solutions using gastight Hamilton volumetric syringes. 1.2 ml of aqueous buffer was then added to each tube, and the chloroform was removed by a modified version of the Rapid Solvent Exchange procedure.[5] Samples were sealed under argon, placed in a temperature controlled water bath at 45.0°C, and then slowly cooled (~ 4°C/hour) to the target temperature where they were held for two days before measurement. Probe/lipid ratios were fixed at 3/10,000 for 18:0-DiO and 3/1000 for DHE.

**SP-FRET measurements.** Fluorescence measurements were carried out on a Hitachi F4500 fluorescence spectrophotometer in photometry mode (10.0 sec integration; 5.0/10.0 mm slits) using a temperature-controlled cuvette holder (Quantum Northwest, Inc). For measurements of DHE-excited DiO fluorescence ($F_{DiO}^{DHEex}$), excitation/emission channels were set to 325/505nm. Meticulous background, bleed-through and 'cross-talk' corrections[6] were provided for. In brief, the F4500 was set up to record four channel combinations for each sample: a scattering signal (430/430nm) and three separate fluorescence signals ($F_{DHE}^{DHEex}, F_{DiO}^{DHEex}, F_{DiO}^{DiOex}$). Calibration standards (i.e., probe-free and single-probe samples) were included in every set of measurements, and periodic closed-shutter integrations were collected for dark current correction. After the raw fluorescence data had been corrected for each possible form of background signal (i.e., dark current, scattering and spurious fluorescence), spectral deconvolution was performed, with the calibration standards serving as quality control samples. In order to correct for sample-to-sample variance, the $F_{DiO}^{DHEex}$ signal from each sample was normalized in proportion to its directly excited donor ($F_{DHE}^{DHEex}$) and acceptor ($F_{DiO}^{DiOex}$) fluorescence signals

$$\left(F_{DiO}^{DHEex}\right)_{norm} = \frac{F_{DiO}^{DHEex}}{\sqrt{F_{DHE}^{DHEex} \cdot F_{DiO}^{DiOex}}}$$

before computing its deviation from a maximum (single-phase) reference value:

$$\Delta(F_{DiO}^{DHEex})_{norm} = (F_{DiO}^{DHEex})_{norm} - (F_{DiO}^{DHEex})_{norm}^{max}$$

### Results and Discussion

Figure 1 shows measurements of Dehydroergosterol-excited 18:0-DiO fluorescence ($F_{DiO}^{DHEex}$) plotted vs. lipid composition in DOPC/DPPC/Cholesterol mixtures at 25°C. The left-hand panel is a 3D scatter plot, in which each data point corresponds to an independently prepared and measured sample. The FRET probes used in these experiments, DHE and 18:0-DiO, happen to partition into different phases[†] in each of the three coexisting phase regimes, which causes each regime to be marked by a reduction in $\left(F_{DiO}^{DHEex}\right)_{norm}$. For this reason, the vertical axes in Fig. 1 have been inverted to make it easier to view the regimes. The data in Fig. 1 have also been color-coded by magnitude to aid the eye.

Each of the three room-temperature coexistence regions that are known to be present in DOPC/DPPC/Chol are clearly revealed in the scatter plot of unsmoothed SP-FRET data. The corresponding surface plot (Fig. 1, right panel) simply makes the structure of the scatter plot easier to see in a static, fixed-angle view (see Supporting Information). Dramatic decreases in FRET efficiency—caused by differential partitioning of the probes into opposite, coexisting domains—clearly identify the boundaries of each two-phase coexistence regime.

The left-hand panel in Fig. 2 shows the same 25.0°C data represented as a 2D mosaic plot. Color-coding identifies the same coexistence regions A, B and C. As the temperature is raised to 35°C (center panel), the $L_\alpha$-$L_\beta$ and $L_\alpha$-$L_o$ regimes shift to higher-DPPC compositions, and by 45°C (right panel) no evidence remains of either region B or region C. The region-A boundary, however, does not appear to shift perceptibly between 25.0°C and 45.0°C.

Figure 3 shows the same data sets as 2D contour plots. In order to generate these plots, each 3D data set was fit with a smooth surface (as in Fig. 1), and that surface was then used to generate constant-interval contour lines. The same regions A, B and C appear in these contour plots as in Fig. 2, but more detail is revealed to the eye. Temperature-dependent changes in the shapes and locations of regions B and C are particularly clear.

---

[5] Buboltz, J.T.; Feigenson, G.W. *Biochim. Biophys. Acta* **1999,** 1417, 232-245.

[6] Berney, C.; Danuser, G. *Biophys. J,* **2003,** *84,* 3992-4010.

[†] DHE is a cholesterol analog, so it partitions as cholesterol does. The phase-preference of 18:0-DiO is as follows: $L_\beta > L_\alpha > L_o >$ chol-xtal.

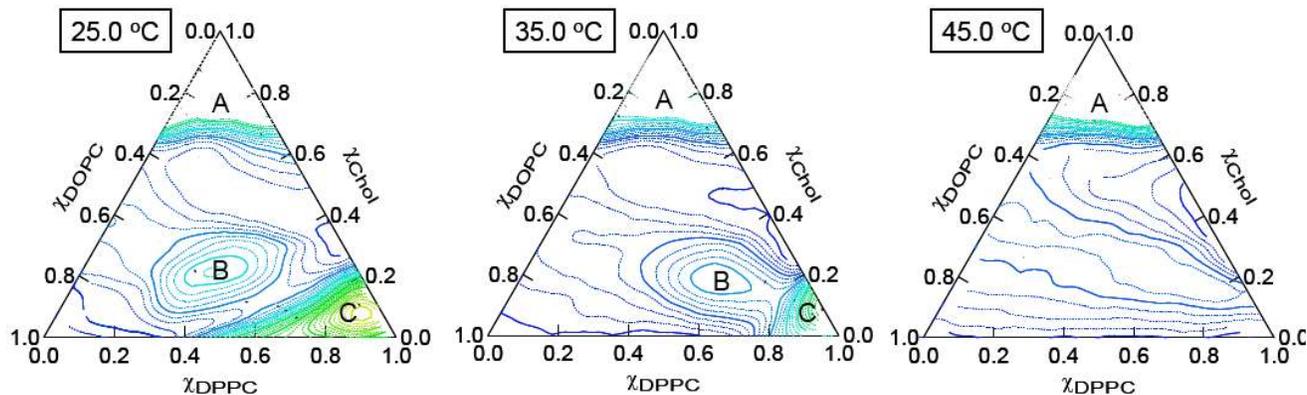

Fig. 3. Effect of temperature on phase boundaries: 2D Contour Plots. The data in Fig. 2 were fit with smooth surfaces, and these were used to generate contour lines at uniform intervals of $\Delta(F_{DiO}^{DHEex})_{norm}$. The same general temperature-dependent phase behavior can be seen here as in as in the mosaic plots, but more detail is revealed by the contour lines.

The results presented in Figs. 1, 2 and 3 provide a compelling demonstration of the general utility of SP-FRET as a tool for mapping lipid-mixture phase behavior. These data show that SP-FRET experiments are: (i) sensitive to a variety of combinations of coexisting lipid phases—even when carried out with a single donor-acceptor probe pair; (ii) can be easily adapted for high-sample-count experiments—the number of independently prepared samples in these experiments was 1294; and (iii) are readily employed in studies of temperature dependence. For all these reasons, SP-FRET experiments have the power to map previously obscure features of membrane-lipid phase behavior with surprising clarity.

To illustrate, we note the following features of DOPC/DPPC/Cholesterol phase behavior that are evident in Figs. 2 and 3. First, the region-A boundary appears remarkably flat at $\chi_{chol} \approx 0.67$ (i.e., invariant within +/- 2 mole% cholesterol for all DOPC/DPPC ratios). To our knowledge, this is the first time that this cholesterol-saturation boundary has been thoroughly mapped in a PC/PC/Cholesterol mixture, but it does confirm expectations established in earlier binary-mixture experiments by Huang et al.[7] Second, the fact that the region-A boundary does not change detectably over a temperature range of twenty degrees is further evidence of precipitously steep dependence on composition of cholesterol's chemical potential at this boundary.[8,9] Third, in contrast to region A, the boundaries of regions B and C are conspicuously sensitive to temperature. As temperature increases from 25°C to 35°C, region C moves toward higher DPPC content and shrinks dramatically—as it must, region C being already bounded near the DPPC-corner of composition-space. The same ΔT causes region B to shift toward higher DPPC content, but since all its coexisting phase compositions are free to increase in DPPC, the overall size of region B does not appear to shrink much, if at all, between 25°C and 35°C. Fourth, inspection of the region-C fluidus boundary at 25°C reveals that a wide range of different fluid-phase compositions can coexist with the gel phase near room temperature, while at 35°C, the fluid-phase compositions which can coexist with gel are constrained very closely to $\chi_{DPPC} \approx 0.80$. And finally, regions B and C appear to be clearly distinct at 25°C, but by 35°C they are either very closely apposed or else partially in contact.

A more general feature of our results is also evident. Although our phase diagrams are qualitatively consistent with previously published reports for DOPC/DPPC/Cholesterol near room temperature (i.e., in terms of the number of two-phase regimes and the nature of the phases coexisting in each), there are clear quantitative differences that cannot be ignored. For example, one may compare our region-B boundaries with the liquid-liquid coexistence boundaries estimated by CFM. According to our data, region B has a high-cholesterol boundary that does not exceed $\chi_{chol} \approx 0.33$, but CFM studies of DOPC/DPPC/Cholesterol report coexisting liquid phases all the way up to $\chi_{chol} \approx 0.50$.[10] This is a significant difference in composition—more than ten times the range of experimental uncertainty—but the discrepancies do not end there. Whereas our observations indicate that just a ten-degree temperature increase causes the entire $L_\alpha$-$L_o$ region (i.e., all boundary compositions) to shift to higher DPPC-content (Fig. 3a cf. 3b), CFM experiments report temperature-dependent shifting in the DOPC-rich $L_\alpha$ boundary only. In stark contrast to our observations, CFM studies report that the DOPC-poor $L_o$ boundary appears perfectly static between 10°C and 50°C.[10,11]

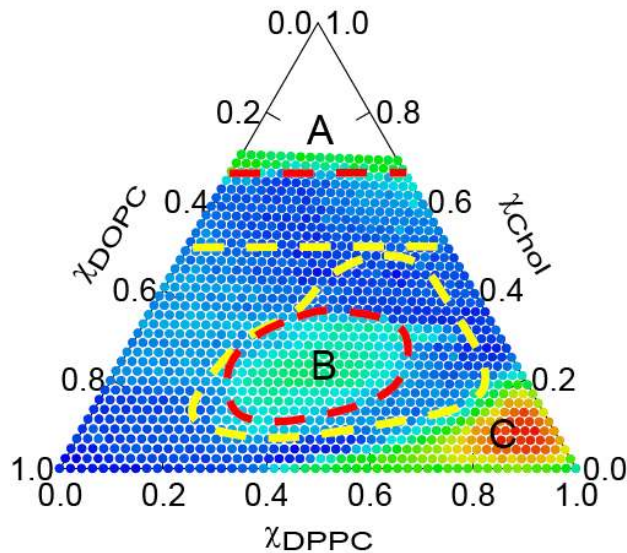

Fig. 4. Studies of lipid-raft mixtures based on conventional sample-prep procedures yield larger apparent regimes of immiscibility. Shown above for comparison are phase boundaries taken from studies based on either rapid solvent exchange (red-dashed lines) or film-deposition (yellow-dashed lines) procedures. The 25°C, region-B miscibility gap reported in this work (red-dashed oval) is considerably smaller than the $L_\alpha$-$L_o$ region estimated by CFM (yellow-dashed pear shape).[11] The same can be said for the saturating cholesterol concentration defining region A (red-dashed horizontal line) and that implied by binary-mixture studies based on film deposition (yellow-dashed horizontal line).[7,13]

---

[7] Huang, J; Buboltz, J.T.; Feigenson, G.W. *Biochim. Biophys. Acta* **1999**, 1417, 89-96.

[8] Huang, J.; G.W. Feigenson. *Biophys. J.* **1999**, 76, 2142-2157.

[9] Ali, M. R.; Cheng, K. H.; Huang, J. *Proc. Natl. Acad. Sci. USA* **2007**, 104, 5372-5377.

Instead of attempting a more detailed comparison of our results with the rather extensive body of published work on DOPC/DPPC/Cholesterol, we would like to focus on the fact that the differences between our present study and those previously published—including most non-CFM studies—are not limited simply to the investigative techniques used. An important difference is, in fact, the method of sample preparation.

As a matter of convenience, conventional model-membrane sample-prep procedures entail depositing a solvent-free lipid film as an intermediate step. Moreover, certain investigative techniques (e.g., CFM or solid-state NMR[12]) require film deposition during sample preparation in order to produce membranes with particular properties (i.e., GUVs or oriented bilayers). It is therefore not surprising that the vast majority of raft-mixture studies have employed film-deposition procedures of one form or another. But it is also quite possible, unfortunately, that the process of film deposition may favor artifactual demixing of sample components, especially in cholesterol-rich mixtures.[7]

Unlike the majority of previous raft studies, we have prepared all our samples by a procedure specifically designed to minimize artifactual demixing. Because SP-FRET measurements can be carried out on ordinary, polydisperse multilamellar vesicle suspensions, we are able to prepare our samples by rapid solvent exchange (RSE), a procedure that avoids the formation of an intermediate lipid film during sample preparation.[5]

Figure 4 illustrates, for DOPC/DPPC/Cholesterol mixtures at 25°C, the differences between phase boundaries we observe using RSE samples and those reported by others using conventional (film-deposition) samples. Our $L_\alpha$-$L_o$ miscibility gap (red-dashed oval) is considerably smaller than the miscibility gap reported by CFM (yellow-dashed pear shape).[11] Likewise, our PC-cholesterol miscibility limit (red-dashed horizontal line) is considerably higher than the limit reported by most workers (yellow-dashed horizontal line) who study PC samples prepared by conventional techniques involving film deposition.[7,13] We hasten to add that we have in fact tested the stability of our smaller miscibility gaps: After completing the temperature-dependent measurements shown in Fig. 2, we incubated (sealed under argon in the dark) all the SP-FRET samples for 60 days at room temperature. We then re-measured all 1294 samples at 25°C. Except for a slight degradation in the quality of the $\left(F_{DiO}^{DHEex}\right)_{norm}$ signal (data not shown), absolutely no shift in any of the regions A, B or C boundaries—all of which were still quite clearly resolved—was observed.

The comparisons in Fig. 4 are clearly suggestive, given that film-deposition procedures can reasonably be expected to favor demixing of lipid-raft systems. During conventional preparation of model membranes, the solvent-free lipid film constitutes an "intermediary solid state," and as such increases the likelihood of artifactual separation of dissimilar mixture components.[5] This effect is of particular concern with cholesterol-rich mixtures, of which lipid-raft systems are a subset. While we certainly do not mean to suggest that the many excellent, previously published studies of raft-mixture phase behavior should be set aside, we do think it is important to take very seriously the possible effects that sample preparation may have had on the apparent phase behavior reported by those studies.

Indeed, we wish to stress that we are not the first group to warn that sample preparation procedures may influence the apparent phase behavior of lipid-raft mixtures. In their 2005 review of lipid-raft studies,[2] for example, Veatch and Keller discussed the difficulties inherent in trying to prepare compositionally uniform GUVs. And Silvius, in his 2003 review,[14] noted more generally the challenges associated with preparing homogeneous phospholipid-cholesterol model membranes, especially at higher cholesterol concentrations.

## Conclusions

In this letter, we have presented high-resolution SP-FRET results for DOPC/DPPC/Cholesterol model membranes between 25°C and 45°C. Our results provide a rich description of the phase behavior exhibited by this canonical lipid-raft mixture, but they also suggest that the apparent phase behavior of lipid-raft mixtures may be particularly susceptible to artifactual demixing during conventional sample preparation procedures. Since, in addition to RSE, there are a variety of other procedures[15,16,17,18] which also avoid the deposition of an intermediary solid-state film, it is our hope that model-membrane researchers will make greater use of alternative sample-prep procedures in future studies of lipid-raft phase behavior.

**Acknowledgment.**
This work was supported by Research Corporation Award CC6814.

**Supporting Information Available:** A rotating animation of the 3D scatter plot in Fig. 1 is available for viewing online at http://pubs.acs.org.